\documentclass[acmsmall]{acmart}
\usepackage[ruled,linesnumbered]{algorithm2e}

\AtBeginDocument{%
  }

\begin{document}

\title{Efficient and Accurate Image Provenance Analysis: A Scalable Pipeline for Large-scale Images}


\author{Jiewei Lai}
\affiliation{%
  \institution{University of Science and Technology of China,Hefei, China}
  \city{Hefei}
  \country{China}}
\email{jw_lai@mail.ustc.edu.cn}

\author{Lan zhang}
\affiliation{%
  \institution{University of Science and Technology of China,Hefei, China}
  \city{Hefei}
  \country{China}}
\email{zhanglan@ustc.edu.cn}

\author{Chen Tang}
\affiliation{%
  \institution{University of Science and Technology of China,Hefei, China}
  \city{Hefei}
  \country{China}}
\email{chentang1999@mail.ustc.edu.cn}

\author{Pengcheng Sun}
\affiliation{%
  \institution{University of Science and Technology of China,Hefei, China}
  \city{Hefei}
  \country{China}}
\email{speical0806@mail.ustc.edu.cn}


\begin{abstract}
The rapid proliferation of modified images on social networks that are driven by widely accessible editing tools demands robust forensic tools for digital governance. Image provenance analysis, which filters various query image variants and constructs a directed graph to trace their phylogeny history, has emerged as a critical solution. 
However, existing methods face two fundamental limitations: First, accuracy issues arise from overlooking heavily modified images due to low similarity, while failing to exclude unrelated images and determine modification directions under diverse modification scenarios. Second, scalability bottlenecks stem from pairwise image analysis that incurs quadratic complexity $O(n^2)$, hindering application in large-scale scenarios.
This paper presents a scalable end-to-end pipeline for image provenance analysis that achieves high precision with linear complexity $O(n)$.  
The pipeline improves filtering effectiveness through modification relationship tracing, which enables the comprehensive discovery of image variants regardless of their visual similarity to the query. 
In addition, the proposed pipeline integrates local features matching and compression artifact capturing, enhancing robustness against diverse modifications and enabling accurate analysis of images' relationships. This allows the generation of a directed provenance graph that accurately characterizes the image's phylogeny history.
Furthermore, by optimizing similarity calculations and eliminating redundant pairwise analysis during graph construction, the pipeline achieves a linear time complexity of O($n$), ensuring its scalability for large-scale scenarios.
Experiments demonstrate our pipeline's superior performance, achieving a 16.7–56.1\% accuracy improvement over existing methods. Notably, it exhibits significant scalability, with an average 3.0-second response time on 10 million-scale images, which is far shorter than the state-of-the-art approach's 12-minute duration.

\end{abstract}




\maketitle

\section{Introduction}
With the significant advancements in image editing technology and the rapid proliferation of social media platforms, digital images can be swiftly modified into numerous variants and disseminated widely within a short span of time. This phenomenon  has fostered a series of illegal activities, including the spread of false images and copyright infringement\cite{DBLP:journals/tkdd/HuangGSLWZ24,DBLP:journals/tkde/LiangTHZZ23,DBLP:journals/tkde/FengYLHLLO25,10.1145/3722225}.
For instance, a fraudulent image depicting an explosion near the Pentagon was posted on Twitter, leading to its widespread circulation and a significant decline in the S\&P 500 and Dow Jones stock market indices.
To mitigate these negative impacts, many researchers are conducting extensive studies for harmful content analysis and false image detection\cite{10.1145/3713077,DBLP:journals/tkde/LiangTHZZ23,DBLP:journals/tkde/GaoSCGLW23,DBLP:journals/tkde/FengYLHLLO25,10.1145/3722225}.
However, merely discovering harmful and false images is not the end. 
Conducting thorough source identification and image modification relationship(s) (MR) analysis provides essential insights for internet governance, facilitating more precise accountability determination \cite{9316916,DBLP:conf/aaai/YangHCL022,zhang2024image,DBLP:conf/icml/LongpreMOBSGPK24}.

\begin{figure}[t]
  \centering
  \includegraphics[width=0.8\linewidth]{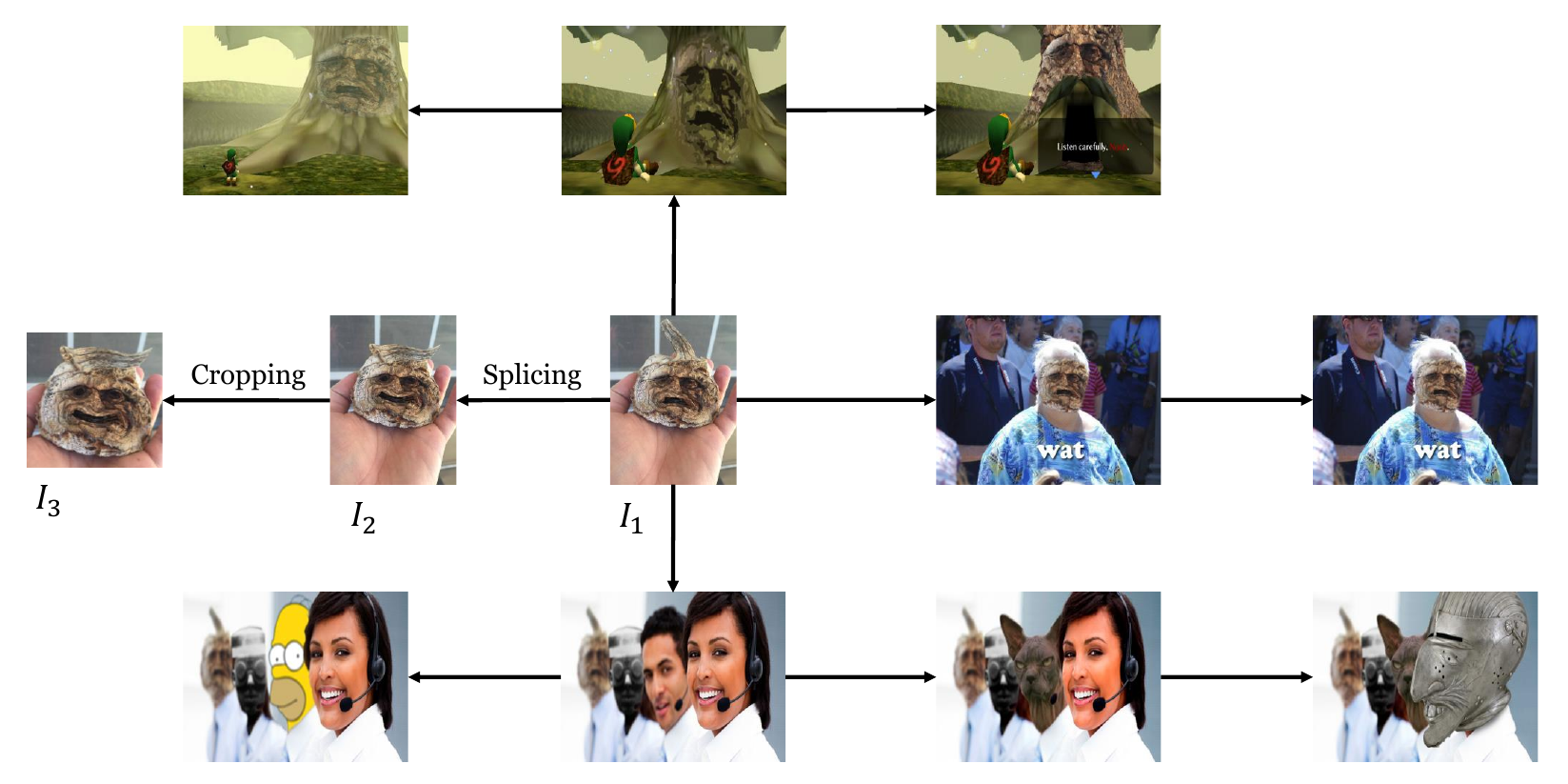}
  \caption{Example of a provenance graph from the Reddit dataset, where nodes represent images and directed edges denote modification relationships.}
  \Description{This figure presents an example of a provenance graph. In this graph structure, images are represented as nodes and directed edges connect image pairs to indicate a direct modification relationship (MR). For instance, if image $I_1$ is modified to produce a new image $I_2$, a directed edge is drawn from $I_1$ to $I_2$, with the direction signifying that $I_2$ is derived from $I_1$.}
  \label{fig:graph_example}
\end{figure}

This motivates research in image provenance analysis, a promising approach \cite{8296535,9316916,8438504,zhang2024image} that can (1) \textit{Provenance Filtering}: discover a set of images that potentially share MR with the query image from a large-scale image database and (2) \textit{Provenance Graph Construction}: this involves mining the relationships among images and generate a directed acyclic graph. In this graph structure, each image is represented as a node and the edges connecting these images indicate the MR between images. The direction of an edge signifies the modification direction, showing that one image is modified from another. See Fig.\ref{fig:graph_example} for an illustrative example.
This technology provides insight into the sources, history and relationships of image data, serves as an essential tool in forensic investigation \cite{DBLP:journals/tkde/MuPZ24} and a powerful framework for academic research in other fields\cite{8438504}.

Recent studies in image provenance analysis \cite{8296535,9316916,8438504,zhang2024image} have demonstrated initial successes. 
These approaches primarily rely on image similarity to discover images that potentially share MR with the query image during provenance filtering. While effective for simple modifications, struggles with images undergoing excessive or large granularity modifications. Such images often exhibit low similarity to the source image, leading to their exclusion from the discovering results.
During the construction of the provenance graph, existing methods typically employ the minimum spanning tree algorithm (MST) \cite{kruskal1956shortest} to generate the undirected provenance graph based on an adjacency matrix derived from a dissimilarity matrix. Though computationally efficient for connecting similar images, MST-based approaches exhibit limitations in two key aspects: they lack robustness as they fail to exclude images without MR, resulting in the inclusion of unrelated images in the provenance graph.
Additionally, the undirected nature of edges struggles with accurately specifying modification directions. While some studies attempt to address this through direction determination \cite{8438504,zhang2024image}, diverse and complex modification techniques (e.g. copy-cove, splicing), and multi-stage modification sequences still hinder precise direction determination, limiting overall accuracy.
Regarding computational efficiency, existing methods employ local features to calculate image similarity through descriptors-wise or patch-wise calculation, which enables fine-grained analysis between image pairs during filtering and graph construction phases.  Though approaches like Optimized Product Quantization \cite{DBLP:conf/cvpr/GeHK013} mitigate some bottlenecks during filtering phases \cite{8296535}, the inherent $O(n^2)$  complexity remains a critical barrier to scalability, particularly as the number of images grows. Furthermore, there is still a lack of research on effectively optimizing the computational efficiency of graph construction. 
Therefore, enhancing the overall effectiveness and scalability of these techniques, particularly for large-scale datasets, necessitates substantial improvements in both provenance analysis accuracy and computational overhead reduction.

In this paper, we aim to design an accurate and efficient image provenance analysis pipeline tailored for large-scale scenarios. However, several critical challenges, exacerbated by diverse and complex modifications and the necessity to process massive image datasets, hinder both provenance accuracy and computational efficiency. 
From the perspective of provenance accuracy, as mentioned earlier, diverse and complex modifications can result in images that exhibit low similarity to the original,  which complicates the task of discovering such images through provenance filtering.
Furthermore, although each modification technique leaves unique traces within the image, the image may undergo multiple rounds of modifications, which can overlap and obscure these traces.  This makes it challenging to analyze the sequence of modifications applied to an image and determine the direction of modification between images, thereby affecting the accuracy of provenance graph construction.
image provenance analysis necessitates conducting provenance filtering across massive image sets and analyzing relationships among multiple image pairs. The potential existence of multiple images contributing to a modified image further increases the computational overhead of relationship mining. 
These challenges collectively hinder the scalability and reliability of image provenance analysis.

This paper proposes a scalable end-to-end pipeline for image provenance analysis, delivering high-precision results with linear computational efficiency and suitability for large-scale applications.
To enhance the provenance accuracy, in the provenance filtering stage, beyond conventional filtering approaches that rely solely on similarity, we design an MR tracing step dedicated to discovering these images that share MR with the query image but exhibit low similarity. This enhanced step is applicable to all provenance filtering methods, thereby enabling their ability to discover images regardless of their similarity level. 
During  provenance graph construction, we propose an accurate and robust framework.  Our framework first conducts shared content mining to analyze images' relationships by local feature matching. This enhances its robustness to handle diverse modified images and ability to accurately establish undirected edges between image pairs sharing direct MR. Subsequently, it determines precise modification directions through capturing JPEG compression artifacts within images.  This is grounded in our analysis that a modified image contains dual JPEG compression artifacts, whereas its source image only exhibits the initial artifact.
To achieve scalability, we extend our proposed filtering and graph construction approaches into an end-to-end pipeline for large scale. The key improvements include optimizing the similarity computational costs through adapting global image representation, avoiding repetitive and redundant pairwise analysis during graph construction, our pipeline is optimized to analyze solely the relationships between the query image and its candidates. The relationships among candidates are maintained within the database with negligible storage overhead and can be obtained through MR tracing. These measures enable our pipeline to operate with linear time computational complexity per analysis.

The contributions of our work are concluded below:  
\begin{itemize}

\item We propose an end-to-end pipeline for scalable image provenance analysis that tightly couples two key components: image relationship analysis and MR tracing within a unified framework, reducing the time complexity from $O(n^2)$ to $O(n)$, enabling accurate phylogeny history tracing at the ten-million scale.

\item We propose an accurate directed graph construction framework for effectively analyzing image relationships by mining shared content and capturing JPEG compression artifacts, enabling robustly identifying modification relationships and determining modification directions across more than 20 diverse modification scenarios. 

\item We propose an MR tracing mechanism to enhance the filtering accuracy and avoid redundant pairwise relationship analysis by tracing MR maintained within the database. It enables the discovery of images sharing MR with the query image but with low visual similarity while allowing our pipeline to focus on analyzing relationships only between the query image and its candidates, significantly improving computational efficiency.
    
\item  Extensive experiments highlight the superior performance and scalability of our method. It surpasses state-of-the-art (SOTA) techniques, achieving an accuracy improvement of 16.7–56.1\% in end-to-end provenance analysis, and requires only an average of 3.0 seconds per analysis at the ten-million scale, nearly 240 times faster than the 12-minute required by SOTA methods.

\end{itemize}

\section{Related works}
The rapid proliferation of digital images in modern society has necessitated robust solutions for image provenance analysis and image attribution, which are critical for addressing challenges such as misinformation, copyright infringement, and forensic investigations. Image attribution aims to identify the source or creator of an image despite modifications, while image provenance analysis extends this by focusing on tracing the origin, modification history, and relationships between images.
This section reviews the methodologies and recent developments in  both domains.

\subsection{Image provenance}

The investigation of the relationship analysis between image pairs is initially proposed by \cite{10.1145/1459359.1459406} as visual migration maps, which sparks significant interest and inspires a series of further research studies, such as those on multimedia phylogeny\cite{10.1117/12.840235,5711452}. 
Early works primarily focus on scenarios involving a single-parent image,  and the issue of multiple parents is considered by more recent studies \cite{7026082,7305756,DIAS2013178}. In these approaches, phylogeny trees or forests are constructed to represent the MR between an image and its parents. 

Recent developments in image provenance analysis have expanded the scope to large-scale image scenarios, aiming to discover images that share MR with the query image from a large set of images, mine and reconstruct their provenance graphs \cite{8296535,8438504,9316916,zhang2024image}.
In research \cite{8296535,8438504}, images are represented as local descriptors extracted via Speeded-Up Robust Features (SURF)\cite{DBLP:conf/eccv/BayTG06}. Following this, these descriptors are then used to calculate the similarity between images, and provenance filtering is carried out based on this similarity. Subsequently, MST is employed to generate undirected provenance graphs, which connects all images relies on similarity.
Research \cite{9316916,zhang2024image} assumes images that share MR have been discovered, focusing instead on the construction of provenance graphs to more accurately mine MR within image pairs. 
Specifically, a manipulation-aware model is trained using quadruplet loss to enhance the feature extractor's sensitivity to capture slight differences among images \cite{9316916}. 
Similarly, a patch attention mechanism is designed to extract detailed local and global features, and a weighted graph distance loss is incorporated to effectively identify and analyze modified regions within images\cite{zhang2024image}.
Advancing beyond this, some methodologies \cite{8438504, zhang2024image} extend these undirected graphs into directed ones, achieved through homography matrices derived from calculated mutual information based matrices or learnable method, thereby providing a more nuanced representation of image relationships.

While these studies demonstrate progressive improvements, as mentioned above, the challenge is still present both in provenance accuracy and efficiency. For example, the fundamental limitations of MST result in the inclusion of unrelated images remaining unaddressed.  Addressing these challenges is crucial to enhance the overall effectiveness and scalability of these techniques, especially considering the growing need to handle large-scale datasets efficiently.

\subsection{Image Attribution}
The objective of image attribution is to discover an image's source  regardless of any modifications it may have undergone. Researchers make significant contributions to image attribution through metadata \cite{aythora2020multi,8658404,rosenthol2020content,DBLP:conf/ecai2/BureacaA24,DBLP:journals/tweb/UmairBLOL24} 
and watermarking\cite{dong2020watermarking,DBLP:conf/mm/MaGHYLJX22,DBLP:conf/aaai/WangMLYFZY24}.
Both metadata-based and watermarking methods aim to establish the origin and authenticity of digital images by embedding information that can be used to trace back to the original image.
Metadata-based methods utilize existing metadata fields to store information, while watermarking techniques embed invisible or visible watermarks directly into the image content itself. 
However, both approaches face limitations, metadata can be easily removed or altered, while watermarks may not always withstand sophisticated attacks or extensive modifications.

To address these limitations, image representation has emerged as a more proactive and comprehensive method for image attribution \cite{9578805,9710809,9856980,DBLP:conf/eccv/SogiST24}. 
The authors in \cite{9578805} suggest that most image tampering occurs at the object-level granularity. Consequently, they propose extracting features related to objects within an image to form a global representation.
Similarly, object-centric representation technologies coupled with pre-trained vision and language models are further studied by \cite{DBLP:conf/eccv/SogiST24}.
In another study \cite{9710809}, a scene graph hash is first leveraged for image attribution under the guidance of a modified SIMCLR loss \cite{pmlr-v119-chen20j}, this method leverages scene graphs to capture the relationships between objects, enhancing the robustness of image attribution. The effect of adversarial examples on image attribution is also considered and addressed in \cite{9856980}. Additionally, the authors of \cite{9622213} highlight that tampered and original areas of an image often undergo different JPEG compression, resulting in compression inconsistency, therefore, this property is used for internal element attribution. 

Although extant studies provide valuable insights, image attribution technologies inadequately address image provenance analysis, which requires discovering all variants and conducting a detailed analysis of their relationships, beyond current capabilities. 


\section{Problem Statement}
Images can be modified into numerous different variants via various modification operations, and image provenance analysis aims to discover such images from a large-scale database (with 10 million images) and then construct a directed graph to reveal the image's phylogeny history, including its source, different versions and corresponding MR. 
For a better understanding of this paper, we provide some descriptions of concepts below.

\noindent \textbf{\textit{Modification Relationships (MR)}}: 
describes the relationship between an original image $I$ and its modified variants $I'$, where $I' \leftarrow \mathcal{G}(I)$ with $\mathcal{G}=\{g_1,g_2,g_3, \dotsc, g_n \}$ representing a sequence of modification operations applied to $I$.
Based on whether an intermediate save operation occurs during the modification process, we distinguish two types of MR: 
A \textit{direct MR} indicates there is no intermediate saving operation during modification, while an \textit{indirect MR} involves at least one intermediate save. 
For example, in Fig. \ref{fig:graph_example},  image $I_1$ and image $I_1$ share a direct MR while image $I_1$ and image $I_2$ exhibit an indirect MR.

\noindent \textbf{\textit{Provenance Filtering}} :
aims to discover a set of images that potentially  share direct or indirect MR with the query image $q$ from a large-scale image set $\mathbf{D}$. This goal can be formalized as follows:
\begin{equation}
    \underset{\mathbf{D}_{candidate} \subseteq \mathbf{D}}{\arg\max} \left( |\mathbf{D}_{candidate} \cap \mathbf{\tilde{D}_{MR}}|\right) 
    s.t. 
    \tilde{\mathbf{D}}_{MR} =\{y\in \mathbf{D}_{MR}\setminus \{q\}\}
\end{equation}
where $\mathbf{D}_{MR}$ denotes the collection of source image $I$ and its variants, 
$\mathbf{\tilde{D}}_{MR}$ represents the subset of $\mathbf{D}_{MR}$, explicitly excluding $q$ itself.
And $\mathbf{D}_{candidate}$ indicates the filtering result obtained by applying provenance filtering to $\mathbf{D}$ with respect to $q$.

\noindent \textbf{\textit{Provenance Graph}} :
describes the relationships among images within an image set.
The MR among an image set $\mathbf{D}_{MR}$ can be modeled as $\mathbb{G}=(\mathbb{V},\mathbb{E}$), where $\mathbb{V}=\{v_i| i \in [1,|\mathbf{D}_{MR}|] \}$ denotes the set of vertices, each $v_i$ corresponds to an image $I_i$ from the set $\mathbf{D}_{MR}$.
$\mathbb{E}=\{e_i|i \in [1,m]\}$ represents a set of directed edges between the vertices. Specifically, an edge $e(u,v)$ indicates a direct MR, signifying that the image represented by vertex $v$ is modified from the image represented by vertex $u$.

\noindent \textbf{\textit{Provenance Graph Construction}} : 
aims to mine all MR within an image set by pairwise analysis and generate a directed provenance graph. This stage consists of two tasks: \textit{Relationship Analysis} and \textit{Direction Determination}. Specifically, given an image set $\mathbf{D}_{candidate}$ that is returned by provenance filtering, the task of relationship analysis is to analyze whether an MR exists between pairs of images. An undirected edge is added between images that share a direct MR, while images that do not share any MR with any other images within $\mathbf{D}_{candidate}$ are discarded, thereby constructing the undirected base graph. The task of direction determination determines the direction of these edges, converting the undirected graph into a directed graph.

\section{Methodology}

\subsection{Overview}
In the following sections, we detail our enhanced filtering method and graph construction framework. Finally, we integrate these components into a scalable pipeline for large-scale scenarios.

Our enhanced provenance filtering method focuses on comprehensively discovering these images that share MR with the query image. The insight behind this approach is that images within the database are no isolated entities but exhibit mutual relationships through an MR. 
Based on this principle, our core design is to trace these MR to discover images with low similarity to the query image, which are often overlooked by top-$k$ querying.
The proposed directed provenance graph construction framework employs a hierarchical manner that integrates two key models: an MR analytical model and a direction determination model, which cooperatively generate directed graphs. The MR analytical model is grounded in the intuition that images sharing MR should exhibit the same content. To this end, it analyzes image pairs' relationships through extracting their high-level semantic features and fine-grained local details for feature matching. Subsequently, our analysis reveals that a modified image contains dual JPEG compression artifacts, whereas its source image only exhibits the initial artifact. Motivated by this result, our direction determination model is designed to capture JPEG compression artifacts within images to infer the direction of modification.
Finally, our provenance filtering method and graph construction framework are extended into an end-to-end pipeline for large-scale computational efficiency. 
Specifically, this pipeline reduces the similarity calculation cost by adopting a global image representation and avoids pairwise analysis during graph construction through tracing the MR. These optimizations result in linear time computational complexity and only require an average of 3.0 seconds per analysis on large-scale datasets with up to 10 million entries. 

\subsection{Modification Relationship Tracing for Enhanced Provenance Filtering}

Existing filtering techniques primarily rely on image similarity,
which often results in overlooking images that share an MR with the query image but exhibit low similarity. This prompts us to explore the feasibility of incorporating additional information to overcome this limitation, thus enhancing the provenance filtering.

\begin{figure}[t]
  \centering
  \includegraphics[width=\linewidth]{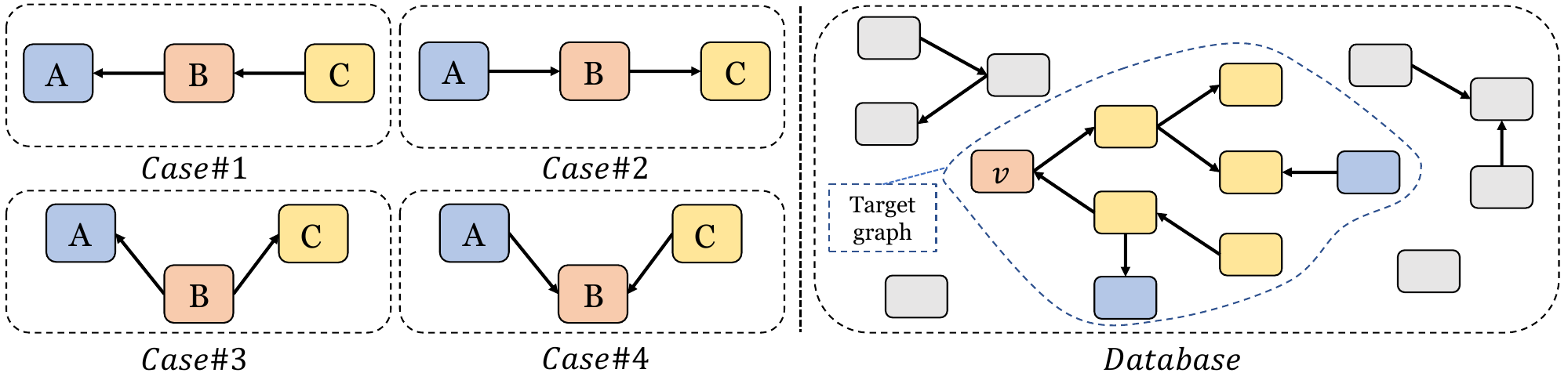}
  \caption{The left panel shows four potential modification relationships for images $I_A,I_B$ and $I_C$. The right panel illustrates an example that shows which images should be returned by MR tracing.}
  \Description{The left panel presents four example provenance graphs for images $I_A$, $I_B$, and $I_C$. Specifically, Case 1 illustrates a sequential modification chain where image $I_C$ is modified into $I_B$, which is further modified into $I_A$, forming the path $I_C \rightarrow I_B \rightarrow I_A$. Case 2 demonstrates a forward modification sequence where $I_A$ is modified into $I_B$, followed by $I_B$ being modified into $I_C$, resulting in the path $I_A \rightarrow I_B \rightarrow I_C$. Case 3 involves a divergent relationship: both images $I_A$ and $I_C$ are derived from a common source image $I_B$, indicated by directed edges from $I_B$ to both $I_A$ and $I_C$. Case 4 represents a convergent modification scenario where images $I_A$ and $I_C$ are independently modified into a shared target image $I_B$, denoted by directed edges from $I_A$ and $I_C$ converging to $I_B$. 
  The right panel illustrates an example demonstrating the images returned by MR tracing. Consider a directed provenance graph where an image $v$ is identified as one of the top-$k$ candidate images. During MR tracing, only images within the graph that share a modification relationship (MR) with $v$ must be returned. This includes both predecessors (images that directly or indirectly modify into $v$) and successors (images modified from $v$), as indicated by the directed edges in the graph. 
}
  \label{fig:filtering}
\end{figure}

\begin{algorithm}[t]
    \SetKwInput{Input}{Input}
    \SetKwInput{Output}{Output}
    \SetKwProg{Fn}{Function}{}{}  
    \SetAlgoLined     
    \Input{Database $\mathbb{D}$, query image $I_q$}
    \text{Extract representation of the query image} $v_q \gets \mathcal{F}_{r}(I_q)$  \\
    \text{Obtain top-$k$ most similar images} $\mathbf{D}_{candidate} = \{I_1, I_2, I_3, \dotsc, I_k\} \gets \mathcal{F}_{f}(\mathbf{D}, v_q, topk)$  \\
    \text{MR tracing} $\mathbb{E} = \{e_1, e_2, \dotsc\} \gets \mathcal{F}_{f}(\mathbb{D}, \mathbf{D}_{candidate})$ \\
    \For{ $e(I_i,I_j) \in \mathbb{E}$ }{
        \If{ $I_i \notin \mathbf{D}_{candidate}$ }{  
            \text{Add overlooked the image to result} \textsc{Add}($\mathbf{D}_{candidate}, I_i$) \\
        }
        \If{ $I_j \notin \mathbf{D}_{candidate}$ }{
            \text{Add overlooked the image to result} \textsc{Add}($\mathbf{D}_{candidate}, I_j$) \\
        }
    }
    \Return $\mathbf{D}_{candidate}$
    \caption{Image Provenance Filtering}
    \label{alg:filtering}
\end{algorithm}

\textbf{Insight.} Traditional filtering methods tend to treat images within a database as isolated entities, disregarding potential MR among them that could be utilized to refine the filtering process. In reality, these MR present an opportunity to improve the discovery of images sharing MR. For example, if an MR $e(I_A, I_B)$  exists between the image $I_A$ and the image $I_B$, conventional methods may fail to identify $I_B$ if it is not among the top-$k$ similarity candidates. However, by tracing the MR associated with the image $I_A$, the image $I_B$ can also be uncovered even when its similarity to the query is insufficient. Although this requires additional information about the MR, it is feasible given that such an MR can be known or pre-analyzed by the database maintainer and subsequently maintained within the graph database. 
Consequently, when an image is identified as one of the top-$k$ most similar candidates to the query image, we can trace the graph in which it is located to return other images that share an MR with it.

\textbf{Modification Relationship Analysis.} Given the complexity and associations within a provenance graph, a critical question arises: should all images within a graph be returned when a node is identified as a top-$k$ candidate?
To address this, we conduct an analysis focusing on three images and their four potential modification relationships, as illustrated in Fig.~\ref{fig:filtering} (left panel). Our goal is to ascertain under what circumstances it is necessary to return the image $I_C$ when querying with image $I_A$.
The specific analysis is outlined below: these cases can be categorized into two groups based on MR existence:

(1) image $I_A$ and image $I_C$ share MR ($Case\#1,Case\#2$): In $Case\#1$, image $I_A$ and image $I_C$ share an indirect MR from $I_C$ to $I_A$, indicating that image $I_C$ may contribute content to image $I_A$. From a forensic and copyright protection perspective, this implies that image $I_A$ potentially infringes upon image $I_C$. Similarly, in $Case\#2$, image $I_C$ may infringe upon image $I_A$, thereby necessitating the return of image $I_C$ in both $Case\#1$ and $Case\#2$ in the stage of provenance filtering.

(2) image $I_A$ and image $I_C$ share no MR ($Case\#3, Case\#4$): In $Case\#3$, despite being part of the same provenance graph and possibly sharing some common content, there is no contribution between image $I_A$ and image $I_C$, meaning there is no infringement. Even if they share common elements, these are contributions from image $I_B$. In $Case\#4$, neither MR nor shared content exists between image $I_A$ and image $I_C$, making it unnecessary to return image $I_C$ during the filtering process.

\textbf{Analysis Conclusion.} The conclusion of our analysis indicates that image $I_C$ should only be returned if image $I_A$ and image $I_C$ share MR, exemplified in the $\textbf{Case\#1}$ and the $\textbf{Case\#2}$. Therefore, only those images that share MR with the query image should be returned.  This not only enhances the precision of provenance filtering, but also ensures that unrelated images are not unnecessarily included in the results.  

\textbf{MR Tracing Design.} Extending this principle to the context of a provenance graph, if a node within the graph is discovered, both its  ancestor and descendant nodes should be returned, which corresponds to  scenarios exemplified in $Case\#1$ and $Case\#2$.  Referring to Fig.\ref{fig:filtering} (right panel), an illustrative example is provided:  if image $v$ is discovered during top-$k$ querying, then all images represented as orange nodes which are either ancestors or descendants of image $v$ through shared MR, should be returned.  Conversely, blue nodes, although residing within the same graph, do not share an MR with image $v$, nor consequently with the query image, and therefore do not need to be returned. 
This enhanced phase is referred to as MR tracing, and it is applicable to all provenance filtering methods.

The methodology for our filtering approach is systematically outlined in Algorithm \ref{alg:filtering}, which integrates a conventional top-$k$ query phase and an enhanced MR tracing phase. (This pseudocode uses redundant steps to make the flow easier to follow, but note that these redundancies are eliminated in the optimized implementation. The same applies to subsequent algorithms in this paper.) The first phase serves as a foundational step that obtains the top-$k$ candidates using any standard filtering technique.
The second phase, MR tracing, constitutes the core innovation of our approach.  It traverses the directed graphs maintained in the database to identify images sharing MR with the query image but excluded by the first phase. The integration of these phases effectively ensures the discovery of images sharing MR with the query image, regardless of high or low similarity, thereby achieving refined and accurate provenance filtering.

\subsection{An Accurate Framework for Provenance Graph Construction}

In this section, we propose a novel framework to construct an accurate directed provenance graph. 
Our framework addresses the limitations inherent in existing methods that rely on MST-based approaches for undirected graph construction and visual content features to infer modification directions. Specifically, our framework employs an MR analytical model to mine MR within image pairs through local feature matching, thereby  establishing undirected edges between image pairs that share direct MR based on the confidence scores of the analytical model, and discarding unrelated images. 
Additionally, by analyzing JPEG compression artifacts in individual images, our framework determines precise modification directions, thus converting the undirected graph into a directed provenance graph.

\begin{figure}[t]
  \centering
  \includegraphics[width=\linewidth]{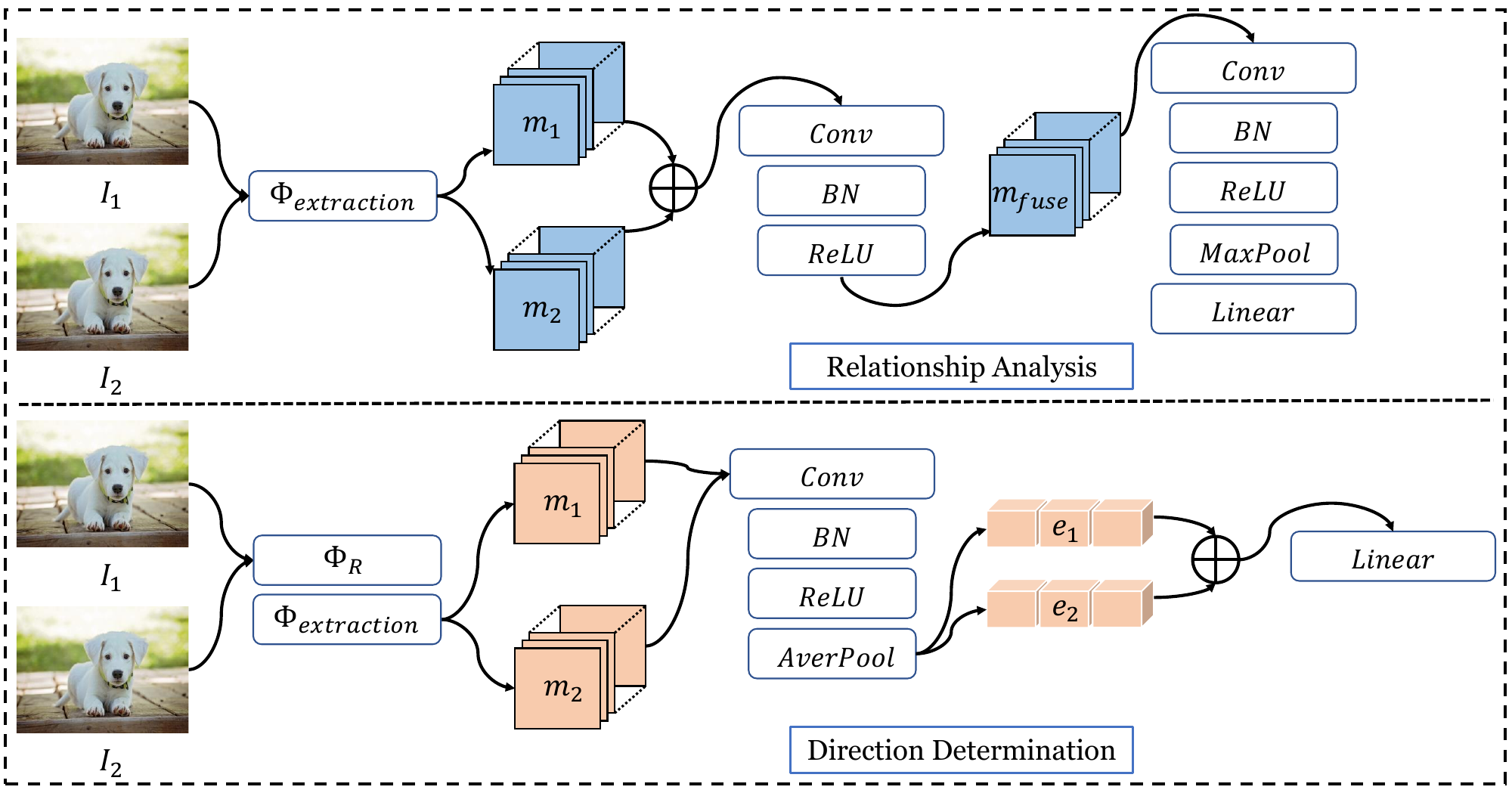}
  \caption{An overview of the proposed framework for constructing directed provenance graphs. The top panel depicts the MR analytical network, while the bottom panel illustrates the modification direction determination network. $\Phi_{R}$ denotes the pre-trained model for reducing compression artifacts, while $\oplus$ represents the channel-wise concatenation operation.}
  \Description{This figure presents an overview of our proposed framework for provenance graph construction and illustrates the architectures of the two core models: the modification relationship analysis model and the direction determination model. Further details regarding the models' design and functionality are provided in the main text.
}
  \label{fig:framework}
\end{figure}

\subsubsection{Local Features Matching for Modification Relationship Mining}

Traditional graph construction methods depend on pairwise dissimilarity metrics and MST algorithms to build undirected graphs. However, these approaches fail to exclude unrelated images remaining after prior provenance filtering, leading to spurious edges connecting image pairs with no actual relational links.
To address this critical limitation, we propose an accurate and robust MR analysis framework, which allows us to establish edges only among image pairs with direct MR and discard unrelated images.

\textbf{Insight.} Images modified from their source often exhibit shared content at multiple levels of detail, with their overlapping regions potentially varying in spatial locations.
Based on this intuition, we design an MR analytical network to analyze whether two images share content through local feature matching, thereby identifying the relationship between them.

\textbf{MR Analytical Network Design. }
The MR analytical network $\mathcal{M}_{undirected}(\cdot)$ is a Siamese network architecture and comprises two primary modules: feature extraction and feature matching (illustrated in Fig.\ref{fig:framework}).
The feature extraction module serves as the backbone for embedding input images into hierarchical feature maps.
It consists of three residual blocks, which preserve both high-level semantic features and fine-grained local details.
For an image pair $\langle I_1, I_2\rangle$ with $I_i \in \mathbb{R}^{CHW}$, where $C, H, W$ denote the channels, height and the width of an image, respectively. (post-uniform resizing and normalization), this module generates corresponding feature maps $\langle m_1,m_2 \rangle, m_i \in \mathbb{R}^{C'H'W'}$ as follows:
\begin{equation}
    m_i=\Phi_{extraction}(I_i;\theta_{extraction})
\end{equation}
where $\Phi_{extraction}(\cdot)$ denotes the feature extraction module with weight parameters $\theta_{extraction}$.
The feature matching module further processes these hierarchical feature maps through three sequential components:
The feature fusion layer generates fused feature maps via  channel-wise concatenation followed by a learnable upsampling convolutional layer:
\begin{equation}
    m_{fuse}=Conv_{upsample}(m_1\oplus m_2)
\end{equation}
where $Conv_{upsample}(\cdot)$ represents a learnable upsample layer and $\oplus$ denotes channel-wise concatenation.
The matching layer learns to mine shared content from fused maps through a dedicated convolutional layer, and the classification head aggregates the output from the matching layer through max-pooling and then makes predictions by a fully connected layer:
\begin{equation}
    \hat{\mathbf{p}}=FC(MaxPool(Conv_{matching}(m_{fuse})))
\end{equation}
The $\hat{\mathbf{p}} \in \mathbb{R}^2$ represents class confidence score.
The complete process of the feature matching module is encapsulated as:
\begin{equation}
    \hat{\mathbf{p}}=\Phi_{matching}(m_1,m_2;\theta_{matching})
\end{equation}
where $\Phi_{matching}$ denotes the matching module with weight parameters $\theta_{matching}$.
The final prediction $\hat{y} \in \{0,1\}$ is determined by comparing class confidence score:
\begin{equation}
\hat{y}=
\left\{
    \begin{aligned}
    &0 \ (\text{MR exist}) ,\  if \ \hat{\mathbf{p}}[0] > \hat{\mathbf{p}}[1] \\
    &1 \ (\text{MR absence}) ,\ otherwise.
    \end{aligned}
\right.
\end{equation}

\subsubsection{Compression Artifacts Capturing for Direction Determination}

Existing direction determination methods rely on pixel-level visual features, but subtle modifications may exhibit minimal perceptual differences, making purely visual-based approaches insufficient for reliable determination. While modification operations often leave distinct forensic artifacts, these traces are frequently destroyed by successive modification operations. Notably, images saved in JPEG format further degrade these modification traces. These limitations highlight the need for exploring robust alternative approaches beyond visual content dependency.

\textbf{Image Modeling and Artifacts Analysis. } JPEG compression, widely used for image storage and transmission, leaves unique artifacts within images that have been extensively exploited in digital forensics for authenticity identification, tampering region localization \cite{10.1145/3474085.3475513,9622213,kwon2022learning}.
Given the pervasive use of JPEG across digital platforms, we explore whether these inherent compression artifacts can be leveraged to determine the direction of MR between image pairs.
Considering a simplified scenario involving two related JPEG images: an original unmodified image and its modified version. Both images are stored using JPEG compression, which introduces characteristic artifacts into the visual data.
An original JPEG image $I_{original}$ that without modification can be modeled as:
\begin{equation}
    I_{original}=I_{original}^0+I_{original}^0 \cdot f_0+\theta 
\end{equation}
where $I_{original}^0$ denotes the ideal uncompressed visual content of the image, $f_0$ denotes the compression artifact induced by the initial JPEG encoding process, and $\theta$ accounts for negligible noise components.
For a modified version created from $I_{original}$ through modification followed by re-compression, we define its representation as:
\begin{equation}
    \begin{aligned}
        I_{modified}&=I_{original}+p+(I_{original}+p) \cdot f_1 +\theta \\
        &=I_{original}^0+I_{original}^0 \cdot f_0 +p + \\
        &(I_{original}^0+I_{original}^0 \cdot f_0 +p) \cdot f_1 +\theta
    \end{aligned}
\end{equation}
Here, $p$ represents the pixel modification introduced by modification operations, while $f_1$ corresponds to new compression artifacts from the second JPEG encoding step.
Let $\mathcal{R}(\cdot)$ denote a learned artifact reduction model capable of estimating and removing JPEG compression artifacts. The residual for the original image $I_{original}$ is then:
\begin{equation}
\begin{aligned}
    r_{orginal}&=I_{original}-\mathcal{R}(I_{original}) \\
    &=I_{original}^0 \cdot f_0
\end{aligned}
\end{equation}
For the modified image, the corresponding residual becomes:
\begin{equation}
\begin{aligned}
    r_{modified}&=I_{modified}-\mathcal{R}(I_{modified}) \\
    &=(I_{original}+p) \cdot f_1\\
    &= I_{original}^0\cdot f_1+I_{original}^0 \cdot f_0 \cdot f_1 +p\cdot f_1
\end{aligned}
\end{equation}

\textbf{Analysis Conclusion.} The key observation lies in the residual structure: $r_{origianl}$ contains only initial compression artifacts ($f_0$), while $r_{modified}$ exhibits dual artifacts through both original and new compression processes($f_0$ and $f_1$).
This structural difference arises from the fact that modified images undergo two JPEG compression processes: the initial compression of the original content and a second compression after modification operations.
By analyzing these residuals, we can distinguish between the inherent compression artifacts of original images and those of modified versions. Specifically, the additional terms in $r_{modified}$ provide directional information about modifications through their dependence on both $f_0, f_1$. This differential analysis enables the conversion of undirected modification edges into directed edges.

\begin{algorithm}[t]
    \SetKwInput{Input}{Input}
    \SetKwInput{Output}{Output}
    \SetKwProg{Fn}{Function}{}{}  
    \SetAlgoLined     
    \Input{Image set $\mathbf{D}_{candidate}$} 
    \text{Initialize adjacency matrix $A$ $\gets$ $N$ $\times$ $N$ zeros}, $N \gets |\mathbf{D}_{candidate}|$  \\
    \For {$I_i \in \mathbf{D}_{candidate}$}{
        \text{Initialize largest confidence} $score_{max} \gets 0$ \\
        \text{Initialize edge index } $index \gets -1$ \\
        \For {$I_j \in \mathbf{D}_{candidate}$}{
            \If {i $\neq$ j}{
                \text{MR mining with MR analytical model} $\hat{p},\hat{y} \gets \mathcal{M}_{undirected}(I_i,I_j)$ \\
                \If {$\hat{y}$=0}{
                    \If{$\hat{p} \geq score_{max}$}{
                        $score_{max} \gets \hat{p}[0]$ \\
                        $index \gets j$
                    }
                }
            }
        }
        \If {A[i][index] = 0}{
        \text{Generate an undirected edge} $A[i][index] \gets$ 1 \\
        \text{Generate directed edge with direction determination model }  $\hat{y}=\mathcal{M}_{directed}(I_i,I_{index})$ \\
        $A[i][index] \gets \hat{y}$ \\
        }
    }
    \text{Construct the directed provenance graph from adjacency matrix} $\mathbb{G(V,E)} \gets A$ \\
    \Return $\mathbb{G(V,E)}$
    \caption{Provenance Graph Construction}
    \label{alg:graph_construction}
\end{algorithm}

\textbf{Direction Determination Network Design. }
To resolve the modification direction between image pairs sharing MR, we propose a dedicated network $\mathcal{M}_{directed}(\cdot)$, which leverages residual artifacts extracted from images to infer directional information and comprises three key modules: feature extraction, artifact representation, and classification.
Given an image pair $\langle I_1, I_2 \rangle$ identified as sharing MR but requiring direction determination, we first extract their residual artifacts using the artifact reduction model $\mathcal{R}(\cdot)$:
\begin{equation}
    r_1=I_1-\mathcal{R}(I_1),
    r_2=I_2-\mathcal{R}(I_2)
\end{equation}
These residuals $r_i$ denote the compression artifacts unique to each image’s encoding history.
The residuals are processed through a feature extraction module $\Phi_{extraction}$, analogous to that in our MR analysis network $\mathcal{M}_{undirected}(\cdot)$. This generates spatial feature maps $\langle m_{r_1},m_{r_2} \rangle,m_{r_i} \in \mathbb{R}^{C'H'W'}$:
\begin{equation}
    m_r=\Phi_{extraction}(r;\theta_{extraction})
\end{equation}
Next, the spatial features map $m_{r_i}$ are projected into global embeddings $\langle e_{r_1},e_{r_2} \rangle, e_{r_i} \in \mathbb{R}^n$ via an artifact representation module ($n$ is the embedding dimensionality), which applies a convolutional layer followed by average pooling:
\begin{equation}
    e_r=AverPool(Conv(m_r))
\end{equation}
Furthermore, the embeddings $\langle e_{r_1},e_{r_2}\rangle$ are concatenated and fed into a classification head for direction prediction:
\begin{equation}
    \mathbf{\hat{p}}=FC(e_{r_1} \oplus e_{r_2})
\end{equation}
where $\oplus$ denotes concatenation and $FC$ represents a fully connected layer outputting class probabilities. 
The final prediction $\hat{y} \in \{-1,1\}$ is determined by comparing class confidence score:
\begin{equation}
\hat{y}=
\left\{
    \begin{aligned}
    &-1 \ (\text{MR direction is from $I_1$ to $I_2$}) ,\  if \ \hat{\mathbf{p}}[0] > \hat{\mathbf{p}}[1] \\
    &1 \ (\text{MR direction is from $I_2$ to $I_1$}) ,\ otherwise.
    \end{aligned}
\right.
\end{equation}

\subsubsection{Hierarchical Modeling of Directed Provenance Graph Construction}

In this section, we present our methodology for directed provenance graph construction using our proposed MR analytical model $\mathcal{M}_{undirected}(\cdot)$ and direction determination model $\mathcal{M}_{directed}(\cdot)$.
These models work in a hierarchical manner to first establish an undirected graph structure through relationship mining, followed by direction determination to specify explicit modification dependencies between images.

The complete workflow of our approach is systematically outlined in Algorithm \ref{alg:graph_construction}. 
Specifically, the undirected MR analytical model is designed to construct the foundational undirected provenance graph. It systematically identifies relationships among all image pairs in a candidate set $D_{candidate}$ (which includes the query image). Images lacking any MR with others are excluded from further processing. For the remaining images, an undirected edge is established between each pair showing the strongest evidence of direct MR interaction (quantified by confidence scores output by $\mathcal{M}_{undirected}(\cdot)$). Following this step, all retained images maintain either direct or indirect MR connections with the query image. These images and relationships are formalized as nodes and edges in an undirected provenance graph, where each edge represents a direct MR between two images.
The direction determination model subsequently refines this undirected graph into a directed graph by determining the direction of the edges. Specifically, image pairs connected via undirected edges are input into $\mathcal{M}_{directed}(\cdot)$ for analysis. This model determines the  modification direction through analyzing the JPEG compression artifacts within images, resulting in a directed provenance graph that explicitly encodes directional relationships between images.

\subsection{A Scalable End-to-End Pipeline for Provenance Analysis}

\begin{algorithm}[t]
    \SetKwInput{Input}{Input}
    \SetKwInput{Output}{Output}
    \SetKwProg{Fn}{Function}{}{}  
    \SetAlgoLined     
    \Input{Database $\mathbb{D}$, query image $I_q$} 
    \text{Extract representation of the query image} $v_q \gets \mathcal{F}_{r}(I_q)$  \\
    \text{Obtain top-$k$ most similar images} $\mathbf{D}_{candidate} = \{I_1, I_2, I_3, \dotsc, I_k\} \gets \mathcal{F}_{f}(\mathbb{D}, v_q, topk)$  \\
    \text{Initialize provenance graph} $\mathbb{G(V,E)} \gets \varnothing$ \\
    \text{Add the query image to vertices set} \textsc{Add}($\mathbb{V}, I_q$) \\
    \For{$I_i \in \mathbf{D}_{candidate}$}{
        \text{MR mining with MR analytical model} $\hat{y} \gets \mathcal{M}_{undirected}(I_q,I_i)$ \\
        \If{$\hat{y}$ = 0}{
            \text{Add candidate image to vertices set} \textsc{Add}($\mathbb{V}, I_i$) \\
            \text{Determine the modification direction} $\hat{y} \gets \mathcal{M}_{directed}((I_q,I_i))$  \\
            \If{$\hat{y}$=-1}{
             \text{Add directed edge to edge set} \textsc{Add}($\mathbb{E},e(I_q,I_i))$ 
            }
            \Else{
            \text{Add directed edge to edge set} \textsc{Add}($\mathbb{E},e(I_i,I_q))$
            }
        }
    }
    \text{MR tracing} $\mathbb{E}_{candidate} = \{e_1, e_2, \dotsc\} \gets \mathcal{F}_{f}(\mathbb{D}, \mathbb{V})$ \\
    \For{ $e(I_i,I_j) \in \mathbb{E}_{candidate}$ }{
        \If{ $e(I_i,I_j) \notin \mathbb{E}$ }{
            \text{Add edge to edges set} \textsc{Add}($\mathbb{E}, e(I_i,I_j)$) \\
            \If{ $I_i \notin \mathbb{V}$ }{  
            \text{Add overlooked image to vertices set} \textsc{Add}($\mathbb{V}, I_i$) \\
        }
            \If{ $I_j \notin \mathbb{V}$ }{
                \text{Add overlooked image to vertices set} \textsc{Add}($\mathbb{V}, I_j$) \\
            }
        }
    }
    \Return $\mathbb{G(V,E)}$
    \caption{End-to-End Image Provenance Analysis Pipeline}
    \label{alg:e2e_pipeline}
\end{algorithm}

The primary computational bottleneck in provenance analysis stems from excessive overhead, severely limiting scalability. This motivates our efforts to optimize computational costs to enable efficient handling of large-scale image datasets. To address this issue, we integrate our proposed provenance filtering and directed graph construction methods into an end-to-end pipeline, forming a scalable framework tailored for large-scale provenance analysis.

The core challenges in large-scale image provenance analysis arise from complex similarity calculations and the analysis of pairwise relationships. 
Firstly, images represented as descriptors \cite{8296535, 8438504} or patch-level features \cite{9316916} require descriptor-wise or patch-wise computations for similarity, resulting in an $O(n^2)$ computational time complexity.
Secondly, constructing a provenance graph demands computing an adjacency matrix based on pairwise analysis, which requires $\frac{n(n-1)}{2}$ calculations and results in an $O(n^2)$ computational time complexity. This further exacerbates computational bottlenecks. Such quadratic computational complexity is impractical for large-scale scenarios.
For example, about 12 minutes are required for a single end-to-end analysis using the method of \cite{8438504}.

To mitigate these challenges, we adopt ISC \cite{yokoo2021contrastive}, a global image representation method, as the foundation of our filtering phase. This enables rapid discovery of top-$k$ candidate images while reducing similarity calculation complexity to $O(1)$. Furthermore, by leveraging these MR maintained within the database, our pipeline is optimized to analyze solely the relationships between the query image and its candidates, thereby avoiding pairwise analysis during the graph construction phase. Rather than redundantly analyzing the relationships among candidate images, we directly obtain them through MR tracing.

Building upon the above optimizations, we develop a scalable image provenance analysis pipeline tailored for large-scale datasets, detailed in Algorithm~\ref{alg:e2e_pipeline}. 
Our full pipeline systematically integrates three stages:
(1) We use ISC to represent the query image and perform provenance filtering to discover top-$k$ candidate images. 
(2) $\mathcal{M}_{undirected}(\cdot)$ establishes undirected edges between the query image and candidates that share MR. The direction of edges is subsequently determined by $\mathcal{M}_{directed}(\cdot)$.
(3) Our pipeline traces the MR within the database to return images that share MR with the query image but are overlooked by the first stage, and the MR among candidate images. These MR and images returned by MR tracing are then added to the provenance graph.
Ultimately, all images and edges jointly constitute a complete directed graph representing the full provenance of the query image. 
Our pipeline utilizes MR tracing to enhance the accuracy of provenance filtering while avoiding pairwise analysis (including both MR analysis and direction determination), thus reducing the computational overhead during graph construction.

\textbf{Complexity Analysis. }
The time computational complexity of the proposed pipeline is an order of magnitude less than that of existing methods. 
The ISC representation for image representation enables $O(1)$ similarity calculations. During undirected graph construction, our approach requires only $n-1$ operations to analyze relationships between the query image and its candidates. And our pipeline limits the determination of edge directions to a maximum of $n-1$ instances.
Comprehensively, the overall time computational complexity for the complete pipeline is $O(n)$, this linear time complexity makes our pipeline more suitable for large-scale scenarios.

\section{Experiments}

\subsection{Dataset}
We use the following datasets in our experiments and we provide a comprehensive overview of the three provenance datasets in Table \ref{tab1}. 
\begin{itemize}
\item   \textit{Media Forensics challenge Dataset}\cite{guan2021user,8638296}:
We utilize provenance datasets released by NIST: NC2017DEV, MFC2018DEV. We extract directed provenance graphs from these journal files, perform data cleaning  and then use the node with the most associations as the query image. 
\item   \textit{Reddit Dataset}\cite{8438504}: This dataset is a more realistic provenance dataset collected from the Reddit community, which is  characterized by a longer and wider graph structure. Due to download failures, we are only able to utilize 171 graphs for evaluation in our experiment(raw 184 graphs).
\item   \textit{PS-Battles Dataset}\cite{heller2018psBattles}: Collected from the Reddit platform, includes 11k original images and 90k hand-crafted tampered images. 
Each original image is associated with a variable number of tampered images, ranging from 0 to 64, with an average of 8 tampered images. 
\end{itemize}

\begin{table}[tb]
\caption{
The detailed information regarding the three provenance datasets. The \textit{Query} and \textit{Database} columns indicate the number of query images/graphs and the database size, respectively. \textit{A-MR}: the average number of images sharing an MR with the query. \textit{A-DIR, A-inDIR}: the average number of images sharing a direct and indirect MR with the query, respectively. 
}
\begin{center}
\begin{tabular}{llllll}
\toprule
Dataset   & Query & Database & A-MR & A-DIR & A-inDIR \\ 
\midrule
NC2017DEV & 378   & 110075   & 18.06      & 4.69     & 13.37       \\
MFC2018DEV  & 159   & 18897    & 39.79      & 12.60    & 27.19       \\
Reddit    & 171   & 9959     & 53.68      & 44.24    & 9.44        \\
\bottomrule
\end{tabular}
\label{tab1}
\end{center}
\end{table}


\subsection{Evaluation Metrics}
For the evaluation of provenance filtering performance, we employ the metric of \textit{R@K}(recall for the top-$k$ images), as suggested in \cite{8438504} for filtering evaluation. In the task of graph construction and end-to-end provenance analysis, we adopt three commonly used metrics introduced by \cite{team2017nimble}: \textit{vertex overlap (VO)}, \textit{edge overlap (EO)} and \textit{vertex and edge overlap (VEO)}. We utilize \textit{EO$^{\diamond}$} and \textit{VEO$^{\diamond}$} to indicate the score of the directed graphs. 
These metrics evaluate the consistency between the constructed provenance graph $\mathbb{G(V,E)}$ and the reference graph $\mathbb{G'(V',E')}$.
The \textit{VO} and \textit{EO} metrics calculate the F1 scores for the predicted nodes and edges respectively, while \textit{VEO} measures the overall graph overlap by combining node and edge overlaps into a single F1 score:
\begin{equation}
    VO=\frac{2 \times |\mathbb{V} \cap \mathbb{V'}|}{|\mathbb{V}|+|\mathbb{V'}|},  
    EO=\frac{2 \times |\mathbb{E} \cap \mathbb{E'}|}{|\mathbb{E}|+|\mathbb{E'}|},
    VEO=\frac{2 \times (|\mathbb{V} \cap \mathbb{V'}|+|\mathbb{E} \cap \mathbb{E'}|)}{|\mathbb{V}|+|\mathbb{V'}|+|\mathbb{E}|+|\mathbb{E'}|}
\end{equation}

\subsection{Implementation Details}

During model training, we employ the Adam optimizer with an initial learning rate of 1e-4 and use a cosine annealing strategy to adjust the learning rate. Specifically, we use a batch size of 128 and conduct a total of 20 epochs of training. 
The model with the best performance on the validation set is saved for further evaluation. All our experiments are conducted on Tesla-P100 GPUs and the Neo4j graph database.

For the MR analytical network training, we randomly selected 70\% of provenance graphs from the provenance dataset as the training set, with the remaining graphs reserved for performance evaluation. Specifically, in the provenance graphs, all image pairs with direct MR were labeled as positive pairs to indicate the presence of an MR. Negative pairs (indicating no MR) were constructed by randomly selecting an equal number of image pairs without MR. 
For the direction determination model, we first pre-trained the model using the PS-Battles dataset and then fine-tuned it on the target provenance dataset. During pre-training, we formed image pairs using 10k original images and their corresponding tampered images (randomly selecting two tampered variants per original image). Subsequently, we applied the Augly tool \cite{papakipos2022augly} to augment the original images, generating an equal number of augmented images paired with their tampered counterparts. The augmentation operations (randomly selected with randomized parameters) followed the configurations described in \cite{papakipos2022augly}. During fine-tuning, aligned with the MR analytical network, 70\% of the dataset was used for training, with the remaining 30\% reserved for validation.
Within the direction determination model, image residuals were computed by subtracting JPEG compression artifacts from the original image. To capture these artifacts, we employed a pre-trained JPEG compression artifact reduction model \cite{DBLP:journals/tip/MaZPT24}.

\subsection{Evaluation of Image Provenance Filtering Task}

\begin{table}[tb]
\caption{Evaluation of provenance filtering methods across three datasets. Symbol $^{\diamond}$ denotes methods enhanced with MR tracing.}
\centering
\begin{tabular}{l|ccc|ccc|ccc}
\toprule
 &\multicolumn{3}{c|}{\textbf{NC2017DEV}}  &\multicolumn{3}{c|}{\textbf{MFC2018DEV}}  &\multicolumn{3}{c}{\textbf{Reddit}} \\ \midrule
\textbf{Method} & R@10 & R@50 & R@100 & R@10 & R@50 & R@100 & R@10 & R@50 & R@100 \\ \midrule
IPA &0.461 &0.735 &0.748 &0.288 &0.864 &0.874 &0.158 &0.540 &0.589 \\
RS  &0.520 &0.754 &0.767 &0.284 &0.809 &0.835 &0.164 &0.325 &0.350  \\
SE  &0.540 &0.773 &0.784 &0.286 &0.840 &0.847 &0.164 &0.315 &0.336 \\
ISC &0.630 &0.863 &0.868 &0.337 &0.901 &0.902 &0.172 &0.406 &0.424  \\ \midrule

IPA$^{\diamond}$ &0.669 &0.874 &0.889 &0.603 &0.932 &0.930 &0.184 &0.585 &0.643 \\ 
RS$^{\diamond}$  &0.624 &0.823 &0.835 &0.452 &0.850 &0.866 &0.197 &0.372 &0.399  \\
SE$^{\diamond}$  &0.642 &0.842 &0.847 &0.449 &0.873 &0.874 &0.187 &0.351 &0.372   \\
ISC$^{\diamond}$ &0.729 &0.885 &0.886 &0.537 &0.907 &0.907 &0.201 &0.441 &0.459 \\ 
\bottomrule
\end{tabular}
\label{tab2}
\end{table}

To validate the enhancement capability of the proposed provenance filtering method, we apply our method to four representative baselines and measure performance improvements across three datasets. Our experiments validate the  enhancement capability of the proposed provenance filtering method, achieving consistent enhancements across all evaluation dimensions.

\subsubsection{Baseline Methods} 
IPA\cite{8296535,8438504}: this method utilizes SURF to embed an image into multiple local descriptors, enabling robust feature representation under geometric transformations.
RS\cite{DBLP:journals/corr/HeZRS15}: a pre-trained ResNet50 is leveraged as the backbone for global feature extraction. It constructs image embedding by directly using final-layer activations without task-specific fine-tuning.
SE\cite{lee2023senet}: a robust retrieval method that integrates visual and geometric structural information via convolutional self-similarity descriptors.
ISC\cite{yokoo2021contrastive}: a contrastive learning-based method trained to maximize inter-class dissimilarity and intra-class similarity. It won first place in the Facebook AI Image Similarity Challenge (descriptor track) and has been widely adopted across domains \cite{DBLP:conf/bigcom/LuoZLWT23,DBLP:conf/sp/JainCCM23,DBLP:conf/aaai/HeHLJYQZYZ23}.

\subsubsection{Evaluation Results} 
Table \ref{tab2} summarizes the performance of baseline methods and their enhanced versions with our proposed MR tracing technique across three datasets. Notably, the enhanced variants (denoted by $^{\diamond}$) consistently outperform their original counterparts across all evaluation dimensions, with the most prominent improvement observed in Recall@10.
This improvement is significant as discovering the most images that share MR with the query image within a small top-$k$ query can significantly reduce unrelated images in the filtering result while preserving filtering efficiency.

Specifically, all enhanced methods achieve 12.8\% average score improvement in R@10 on the NC2017DEV dataset, and the enhancement demonstrates the most prominent gains on IPA$^\diamond$. for instance, it achieving 20.8\% recall score improvement in R@10 (from 0.461 to 0.669).
The MFC2018DEV dataset shows the most dramatic improvements across all methods, achieving a 21.2\% average score improvement in R@10. The enhanced IPA$\diamond$ achieves a remarkable 31.5\% increase in R@10 score and others outperform their base version by over 16.3\%.
While improvements are less pronounced on the Reddit dataset compared to other datasets due to its whimsical and diverse nature, the enhancements still deliver consistent gains across all metrics.

\subsection{Evaluation of Provenance Graph Task}
To evaluate the effectiveness of our graph construction method, 
in this section, we compare our proposed graph construction approach with three advanced graph construction methods in oracle and disturb modes.

\subsubsection{Baseline Methods}
Existing methods uniformly construct undirected provenance relationships by applying MST to dissimilarity matrices derived from pairwise dissimilarity calculations. However, they differ in their approaches to computing dissimilarities and constructing directed graphs. The key distinctions among baseline methods are summarized as follows:
TAE\cite{9316916}: this method is an undirected provenance construct method, which embeds images into multiple patch-level features and calculates dissimilarity through patch-wise feature calculations. 
IAP\cite{8296535,8438504}: this approach extracts local descriptors from images and computes dissimilarity based on descriptor-wise matching. To determine edge direction, IAP estimates the asymmetric mutual information between image pairs by estimating a homography matrix that maps features from one image to another, thereby inferring geometric transformations.
GEVT\cite{zhang2024image}: this leverages patch embeddings generated by a vision-transformer-based model for computing dissimilarities. After forming an undirected graph structure, it models images and undirected edges into a graph structure and employs a graph neural network to predict edge directions. 

\subsubsection{Experimental Settings}
We conduct experiments under two evaluation modes: (1) \textit{Oracle Mode}:  This mode focuses exclusively on graph construction, assuming that the candidate set returned by the filtering stage contains no unrelated images. (2) \textit{Disturb Mode}: Building upon Oracle Mode, this mode introduces approximately 10\% interfering images into each graph, creating a more realistic scenario to evaluate the method's robustness against interference.

\begin{table}[tb]
\caption{Evaluation of Provenance Graph Construction methods across three datasets under oracle and disturb mode. The best and second-best results are marked in \textbf{bold} and \underline{underlined}.}
\centering
\begin{tabular}{lccc|cc|ccc|cc}
\toprule
   &\multicolumn{5}{|c|}{\textbf{Oracle}}  &\multicolumn{5}{c}{\textbf{Disturb}}   \\ \midrule
\textbf{Method} & \multicolumn{1}{|c}{VO} & EO & VEO & EO$^{\diamond}$ & VEO$^{\diamond}$ & VO & EO & VEO & EO$^{\diamond}$ & VEO$^{\diamond}$   \\ \midrule 
&\multicolumn{10}{c}{\textbf{NC2017DEV}} \\  \midrule
TAE &\multicolumn{1}{|c}{\textbf{1.0}} &\textbf{0.568} &\textbf{0.783} & &  &\underline{0.978} &\textbf{0.557} &\textbf{0.767} & &   \\
IPA &\multicolumn{1}{|c}{0.863} &0.226 &0.554 &0.109 &0.496 &0.863 &0.226 &0.553 &0.109 &0.496 \\
GEVT &\multicolumn{1}{|c}{\textbf{1.0}}  &0.445 &\underline{0.715} &\textbf{0.220} &\textbf{0.603} &\underline{0.978} &\underline{0.451} &\underline{0.711} &\textbf{0.239} &\textbf{0.606}  \\
Ours&\multicolumn{1}{|c}{\underline{0.987}} &\underline{0.450} &0.709 &\underline{0.205} &\underline{0.584} &\textbf{0.994} &0.433 &0.704 &\underline{0.197} &\underline{0.583}   \\  \midrule  
&\multicolumn{10}{c}{\textbf{MFC2018DEV}} \\ \midrule
TAE &\multicolumn{1}{|c}{\textbf{1.0}} &0.279 &\underline{0.615} & & &\underline{0.960} &0.269 &0.592 & &   \\
IPA &\multicolumn{1}{|c}{0.941} &0.191 &0.542 &0.060 &0.472 &0.938 &0.190 &0.538 &0.057 &0.470   \\
GEVT &\multicolumn{1}{|c}{\textbf{1.0}} &\underline{0.284} &0.614 &\underline{0.148} &\underline{0.541}  &\underline{0.960} &\underline{0.285} &\underline{0.598} &\underline{0.158} &\underline{0.530}   \\
Ours &\multicolumn{1}{|c}{\underline{0.994}} &\textbf{0.346} &\textbf{0.643} &\textbf{0.199} &\textbf{0.563} &\textbf{0.997} &\textbf{0.373} &\textbf{0.658} &\textbf{0.226} &\textbf{0.579}   \\  \midrule
&\multicolumn{10}{c}{\textbf{Reddit}} \\ \midrule
TAE &\multicolumn{1}{|c}{\textbf{1.0}} &0.178 &\underline{0.593} & & &\underline{0.957} &0.167 &\underline{0.566} & &   \\
IPA &\multicolumn{1}{|c}{0.898} &\underline{0.195} &0.550 &\underline{0.083} &0.496 &0.893 &\underline{0.192} &0.547 &\underline{0.082} &0.492   \\
GEVT &\multicolumn{1}{|c}{\textbf{1.0}} &0.137 &0.570 &\underline{0.083} &\underline{0.543}  &\underline{0.957} &0.131 &0.546 &0.079 &\underline{0.520}   \\
Ours &\multicolumn{1}{|c}{\underline{0.997}} &\textbf{0.327} &\textbf{0.662} &\textbf{0.263} &\textbf{0.630} &\textbf{0.959} &\textbf{0.314} &\textbf{0.636} &\textbf{0.252} &\textbf{0.605}  \\
\bottomrule
\end{tabular}
\label{tab3}
\end{table}

\subsubsection{Evaluation Results}
The detailed evaluation results are presented in Table \ref{tab3}. These results demonstrate our framework’s ability to effectively mine relationships between images, determine modification directions, and robustly handle unrelated images.

\textbf{Overall Performance.}  Our method achieves superior performance in accurately constructing undirected and directed provenance graphs across both oracle and disturb modes, demonstrating its practical usability and robustness. Specifically, our proposed method achieves the best performance across most evaluation metrics on both MFC2018DEV and Reddit datasets under both oracle and disturb modes. Specifically, in the oracle mode, our approach outperforms the second-best method by 9.2\% VEO score improvement on the Reddit dataset. And in directed graph generation (VEO$^{\diamond}$), we achieve an 8.7\% performance gain over competing methods. This advantage is maintained in disturbed conditions where unrelated images are introduced,  showing 7.0\% VEO score improvement in constructing undirected graphs and 8.5\% VEO$^{\diamond}$ score improvement in directed graphs.
On MFC2018DEV, our method continues to improve various metrics even when unrelated images are included. While performance on NC2017DEV is sub-optimal compared to the best-performing methods, it consistently achieves second-best results across most evaluation metrics.

\textbf{Robustness.}  Our approach demonstrates robust performance in handling unrelated images.
Traditional methods like TAE and GEVT, which rely on MST techniques for undirected edge generation, achieve perfect VO scores only under oracle conditions where no unrelated images exist. However, their performance significantly degrades  when disturbed with unrelated images due to the inability to effectively filter them out. In contrast, our method successfully identifies and discards unrelated images, showing
VO scores improvement on MFC2018DEV and NC2017DEV datasets under disturb mode and minimal performance degradation on Reddit compared to second-best methods.
These indicate that our method exhibits robustness and practical usability to work in realistic scenarios, which can effectively discard images without MR during undirected provenance construction.
More detailed robustness evaluation experiments are conducted in Section 4.7.

\textbf{Effectiveness.} Our proposed method demonstrates the capability to accurately mine direct MR between image pairs and establish undirected edges in provenance graphs, while simultaneously determining the precise direction of modifications. 

(1) In the construction of undirected provenance graphs, our approach achieves SOTA EO scores across most datasets under both oracle and disturbed conditions. These results confirm our method's effectiveness in mining MR through local feature matching and constructing undirected graphs based on confidence score. This is a capability lacking in existing techniques relying on MST and dissimilarity matrices. Notably, we outperform the second-best methods by 6.2\% and 8.8\% in EO scores for MFC2018DEV under oracle and disturb mdoes respectively. This performance margin expands significantly on the Reddit dataset, achieving improvements of 13.2\% (oracle) and 12.2\% (disturb), demonstrating robustness across diverse image distributions. 

(2) For directed graph construction, our method achieves superior directional MR determination with consistent performance advantages. It outperforms competing approaches by achieving 5.1\% and 6.8\% improvements in EO$^{\diamond}$ scores on the MFC2018DEV dataset under oracle and disturbed conditions, respectively. This margin further expands to 18.0\% and 17.0\% gains for the Reddit dataset across both evaluation settings. Additionally, our approach exhibits minimal performance degradation when transitioning from undirected to directed graph structures, as quantified by the metric $\frac{EO^{\diamond} -EO}{EO}$.
The method’s efficacy is constrained on the NC2017DEV dataset due to its diverse image formats. These various formats introduce ambiguities in residuals, which is critical for the direction determination of our model, thereby limiting performance. Addressing this limitation to improve the generation of our method constitutes a key focus of future research. 
In general, the high EO$^{\diamond}$ scores coupled with minimal degradation highlight our method’s effectiveness, driven by its ability to precisely capture JPEG compression artifacts.


\subsection{Evaluation of End-to-End Provenance Analysis Task}

In this experiment, we consider the impact of both provenance filtering and provenance graph construction, perform a comprehensive assessment of the entire end-to-end provenance analysis workflow and present the evaluation results of the final directed provenance graph. We further evaluate the scalability of our pipeline in terms of computational and storage overhead on a large-scale dataset with 10 million images. 

\subsubsection{Baseline Methods}
IPA is an end-to-end provenance analysis method that integrates two stages: filtering and graph construction.
Since GEVT focuses solely on provenance graph construction, we adopt the filtering results from ISC to align with our pipeline's methodology. Baseline: the baseline represents an ablation variant of our full pipeline, constructed by excluding the MR tracing step to verify the effectiveness of MR tracing.

\begin{table}[tb]
\caption{Evaluation of End-to-End Provenance Analysis methods across three datasets. The best results is marked in \textbf{bold}.}
\centering
\begin{tabular}{lccc|ccc|ccc}
\toprule
  &\multicolumn{3}{|c|}{\textbf{NC2017DEV}}  &\multicolumn{3}{c|}{\textbf{MFC2018DEV}}  &\multicolumn{3}{c}{\textbf{Reddit}} \\ \midrule
\textbf{Method}  & \multicolumn{1}{|c}{VO}  & EO$^{\diamond}$ & VEO$^{\diamond}$ & VO  & EO$^{\diamond}$ & VEO$^{\diamond}$ & VO  & EO$^{\diamond}$ & VEO$^{\diamond}$   \\ \midrule
&\multicolumn{9}{c}{\textbf{top-10}} \\ \midrule
IPA  &\multicolumn{1}{|c}{0.624}  &0.051 &0.348 &0.445 &0.034 &0.221 &0.233 &0.003 &0.122 \\
GEVT  &\multicolumn{1}{|c}{0.647} &0.155 &0.402 &0.555 &0.104 &0.309 &0.316  &0.053 &0.186 \\
Baseline   &\multicolumn{1}{|c}{0.748} &0.118 &0.441 &0.541  &0.112 &0.307 &0.315 &0.238 &0.278\\
            &\multicolumn{1}{|c}{\textit{+0.241}} &\textit{+0.590} &\textit{+0.398} &\textit{+0.455} &\textit{+0.660} &\textit{+0.563} &\textit{+0.288} &\textit{+0.246} &\textit{+0.267} \\
Ours  &\multicolumn{1}{|c}{\textbf{0.989}}  &\textbf{0.708} &\textbf{0.839} &\textbf{0.996}  &\textbf{0.772} &\textbf{0.870} &\textbf{0.603}  &\textbf{0.484} &\textbf{0.545}\\
\midrule
&\multicolumn{9}{c}{\textbf{top-50}} \\ \midrule
IPA     &\multicolumn{1}{|c}{0.781}  &0.084 &0.441 &0.859 &0.049 &0.427 &0.655 &0.016 &0.346 \\
GEVT    &\multicolumn{1}{|c}{0.394} &0.087 &0.238 &0.660  &0.102 &0.366 &0.434  &0.051 &0.245 \\
Baseline  &\multicolumn{1}{|c}{0.847} &0.293 &0.500 &0.892  &0.184 &0.513 &0.526 &0.401 &0.464 \\
            &\multicolumn{1}{|c}{\textit{+0.095}} &\textit{+0.364} &\textit{+0.285} &\textit{+0.099} &\textit{+0.538} &\textit{+0.316} &\textit{+0.172} &\textit{+0.186} &\textit{+0.179} \\
Ours  &\multicolumn{1}{|c}{\textbf{0.942}}  &\textbf{0.657} &\textbf{0.785} &\textbf{0.991}  &\textbf{0.722} &\textbf{0.829} &\textbf{0.698}  &\textbf{0.587} &\textbf{0.643}\\
\midrule
&\multicolumn{9}{c}{\textbf{top-100}} \\ \midrule
IPA     &\multicolumn{1}{|c}{0.760} &0.069 &0.423 &0.837  &0.036 &0.411  &0.605 &0.016 &0.319\\
GEVT    &\multicolumn{1}{|c}{0.232} &0.051 &0.141 &0.415  &0.062 &0.232 &0.307 &0.036 &0.173  \\
Baseline   &\multicolumn{1}{|c}{0.830} &0.287 &0.490 &0.888 &0.183 &0.511 &0.493 &0.375 &0.435 \\
            &\multicolumn{1}{|c}{\textit{+0.093}} &\textit{+0.358} &\textit{+0.280} &\textit{+0.097} &\textit{+0.535} &\textit{+0.313} &\textit{+0.152} &\textit{+0.175} &\textit{+0.167} \\
Ours   &\multicolumn{1}{|c}{\textbf{0.923}}  &\textbf{0.645} &\textbf{0.770} &\textbf{0.985}  &\textbf{0.718} &\textbf{0.824} &\textbf{0.654}  &\textbf{0.550} &\textbf{0.602}\\
\bottomrule
\end{tabular}
\label{tab4}
\end{table}

\subsubsection{Evaluation Results}
These results demonstrate our pipeline's significant effectiveness and scalability, outperform existing methods with a 16.7–56.1\% improvement in VEO$^{\diamond}$ scores and only require a few seconds for end-to-end analysis, which are orders of magnitude faster than the SOTA methods. 

\textbf{Effectiveness.}
The proposed method demonstrates consistent SOTA performance across all datasets and evaluation metrics, significantly outperforming existing approaches. The performance of different methods for end-to-end provenance analysis across three datasets is summarized in Table~\ref{tab4}. Our approach achieves 28.0–39.8\% improvements in VEO$^{\diamond}$ scores over competing methods on the NC2017DEV dataset, with even more pronounced gains of 31.3–56.1\% observed on the MFC2018DEV benchmark.  Notably, it maintains superior performance even when applied to the most complex and challenging real-world datasets (Reddit), where its critical metrics improve by 16.7–26.7\% compared to other methods.

The MR Tracing step demonstrates strong enhancement effects, enabling our method to significantly outperform baseline approaches across all evaluation metrics. As shown in Table  \ref{tab4}, after enhancing with MR tracing, all VO scores have improved to varying degrees. This indicates that MR Tracing can effectively obtain images that were overlooked during the initial top-$k$ candidate discovery querying due to low similarity to the query image. The notable improvement in EO$^{\diamond}$ scores further demonstrates its ability to accurately and efficiently return the MR among candidate images that are maintained in the database. The incorporation of MR Tracing enhances our method's provenance accuracy and scalability. Our method not only discovers images with MR that were overlooked during the top-$k$ querying due to their low similarity with the query image, but also obtains the MR among candidate images, enabling us to analyze only the relationships between query and candidate images, thus avoiding redundant pairwise analysis.

It is also worth noting that even without MR Tracing enhancement, our approach still achieves advanced performance compared to existing methods. While the VEO$^{\diamond}$ score under top-10 conditions is slightly lower than GEVT's scheme, we attain the best performance across all other metrics with improvements of 6.7–11.8\% over other approaches. This further validates the accuracy and robustness of our directed provenance construction framework.

\begin{figure*}[t]
    \centering
    \includegraphics[width=0.95\textwidth]{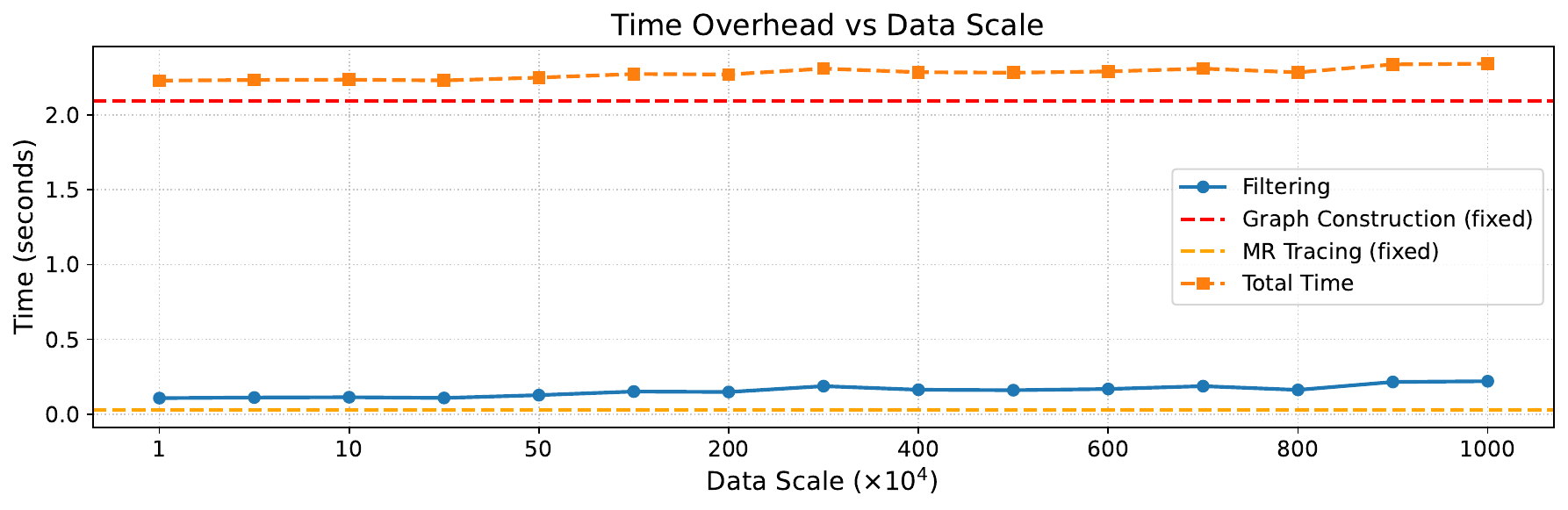}
    \caption{\label{fig:runtime} Time cost of our image provenance analysis pipeline across different data scales.}
\end{figure*}

\textit{\textbf{Scalability.}} 
Our proposed pipeline achieves remarkable computational efficiency with linear computational complexity and only requires negligible storage overhead. 

(1) \textit{Time overhead}: experimental result demonstrates the practical feasibility of our approach in large-scale image provenance analysis tasks. We evaluate the top-$100$ provenance analysis performance of our pipeline across varying data scales, ranging from 10 thousand (10K) to 10 million (10M) image entries, results are detailed in Fig.~\ref {fig:runtime}. The filtering time remains relatively stable throughout the experiments, increasing only marginally with larger data sizes. Specifically, the average query time is 0.107s for a dataset of 10K images and increases to 0.220s when querying 10 million (10M) images. Notably, the total processing time includes both MR tracing (0.027s, fixed) and graph construction (2.096s, fixed), which are independent of the data scale. The overall system maintains efficient performance, with a total per-analysis time of approximately 2.3s even at the 10M scale.  Notably, the IPA method requires more than 12 minutes solely for the graph construction task, which significantly limits its scalability and real-time applicability (We use the results reported in the original paper for comparison, as re-running the method on large-scale datasets becomes computationally prohibitive).


(2) \textit{Storage overhead}: our method effectively reduces the computational complexity from O($n^2$) to linear O($n$), only requiring negligible storage overhead. To quantify the storage overhead, we measure the sizes of all files related to MR in the database. Our analysis reveals that we maintain approximately 12,000 MR, which only requires 0.66 MB (i.e., 680 KB) of storage space.

Our method's efficient time and storage overheads and high scalability make it well-suited for real-world applications requiring rapid and accurate provenance analysis.

\subsection{Quantitative Analysis and Ablation Study }

In this section, we conduct quantitative experiments to evaluate the performance of our MR analytical model and direction determination model across three provenance datasets. Additionally, ablation studies are performed  to validate the effectiveness of pretraining on the PS-Battles dataset for the direction determination model. Finally, we provide the robustness evaluation to demonstrate our models are robust against diverse image modifications.

\begin{figure*}[tb]
  \includegraphics[width=0.75\linewidth]{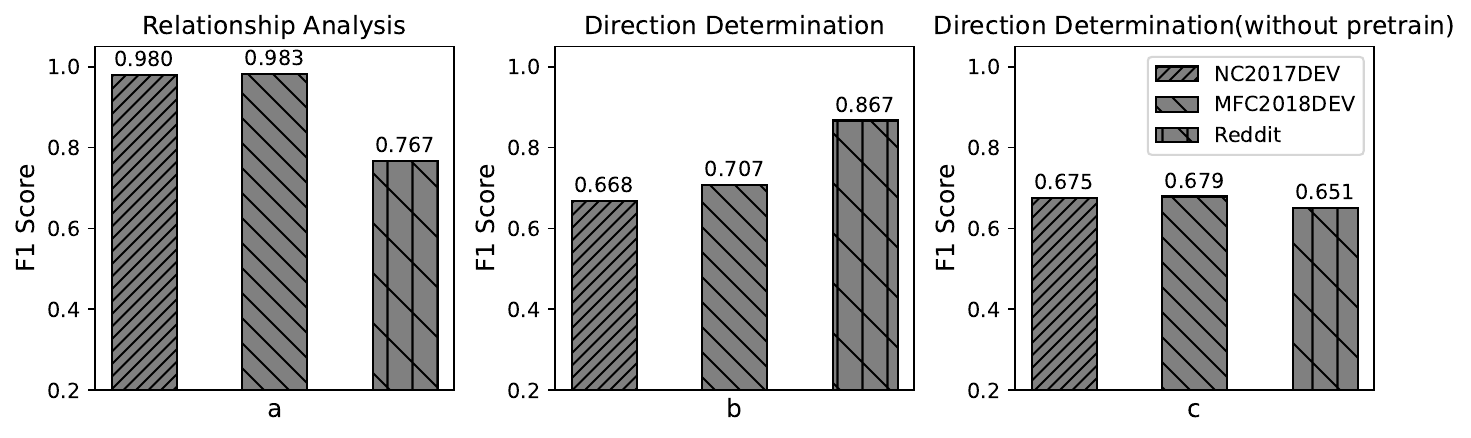}
  \caption{Evaluation of MR analytical and direction determination Models. This figure presents the F1-scores of relationship analysis and direction determination models (with and without pretraining) across three benchmark datasets.
  }
  \Description{
  NC2017DEV: The relationship analysis model achieves 98.0\% F1-score, while the direction determination model (pretrained) attains 66.8\%.
  MFC2018DEV: Relationship analysis achieves 98.3\% F1-score, and the pretrained direction determination model reaches 70.7\%, with a notable drop (20.8\% reduction) in the non-pretrained version.
  Reddit: The direction determination model excels with 86.7\% F1-score, but the non-pretrained variant performs poorly (65.1\%). Relationship analysis shows lower performance (76.7\%) due to complex inter-image relationships.
  }
  \label{fig:f1_score_comparison}
\end{figure*}

\subsubsection{MR Analysis Performance Quantification.}  The proposed MR analytical model demonstrates exceptional performance across benchmark datasets. As illustrated  in Fig. \ref{fig:f1_score_comparison} (a), our models achieve 98.0\% and 98.3\% F1-scores on the NC2017DEV and MFC2018DEV datasets respectively. Notably, the model maintains significant effectiveness, with an F1-score of 76.7\% on the Reddit dataset despite its complex inter-image relationships and diverse modification patterns. These results validate models' capability to mine modification relationships across varying complexity levels effectively.

\subsubsection{Direction determination Performance Quantification.} The proposed direction determination models demonstrate considerable performance across diverse image formats, maintaining effective directional inference even with non-JPEG images. As illustrated  in Fig. \ref{fig:f1_score_comparison} (b), our model achieves an F1-score of 86.7\% on the Reddit dataset, which is the highest among all evaluated benchmarks. This superior performance stems from the Reddit dataset contains predominantly JPEG-compressed images with minor PNG components, aligning well with the model's design that leverages JPEG compression artifacts for directional inference. However, performance decreases on NC2017DEV and MFC2018DEV, which contain diverse image formats (including JPEG, PNG, BMP, TIF, CR2, IIQ, NEF, and MPO). Notably, NC2017DEV exhibits extreme format heterogeneity where about 80.0\% of related image pairs contain non-JPEG formats, even with more than 34.0\% image pairs involving entirely non-JPEG formats. This diverse format distribution limits artifact-based analysis, yet our method maintains effectiveness by achieving 66.8\% and 70.7\% F1-score on NC2017DEV  and MFC2018DEV datasets.

\subsubsection{Ablation Study: Pretraining Efficacy.} We conduct ablation experiments to show the effectiveness of pretraining. We directly train the direction determination models on the provenance datasets without pretraining on the PS-Battles dataset. As illustrated in Fig. \ref{fig:f1_score_comparison} (b and c), the pretraining step enhances model performance by 2.80\% on the MFC2018DEV dataset and 21.6\% on the Reddit dataset in F1-score improvement. The minimal performance variation observed on NC2017DEV is due to its diverse formats. These demonstrate the critical importance of the pretraining step for the direction determination task. 


\subsubsection{Robustness Evaluation.} To evaluate the robustness of the MR Analytical model and direction determination model under diverse image modification scenarios, we test their performance across 22 common modification types.





\begin{figure*}[tb]
  \includegraphics[width=0.95\linewidth]{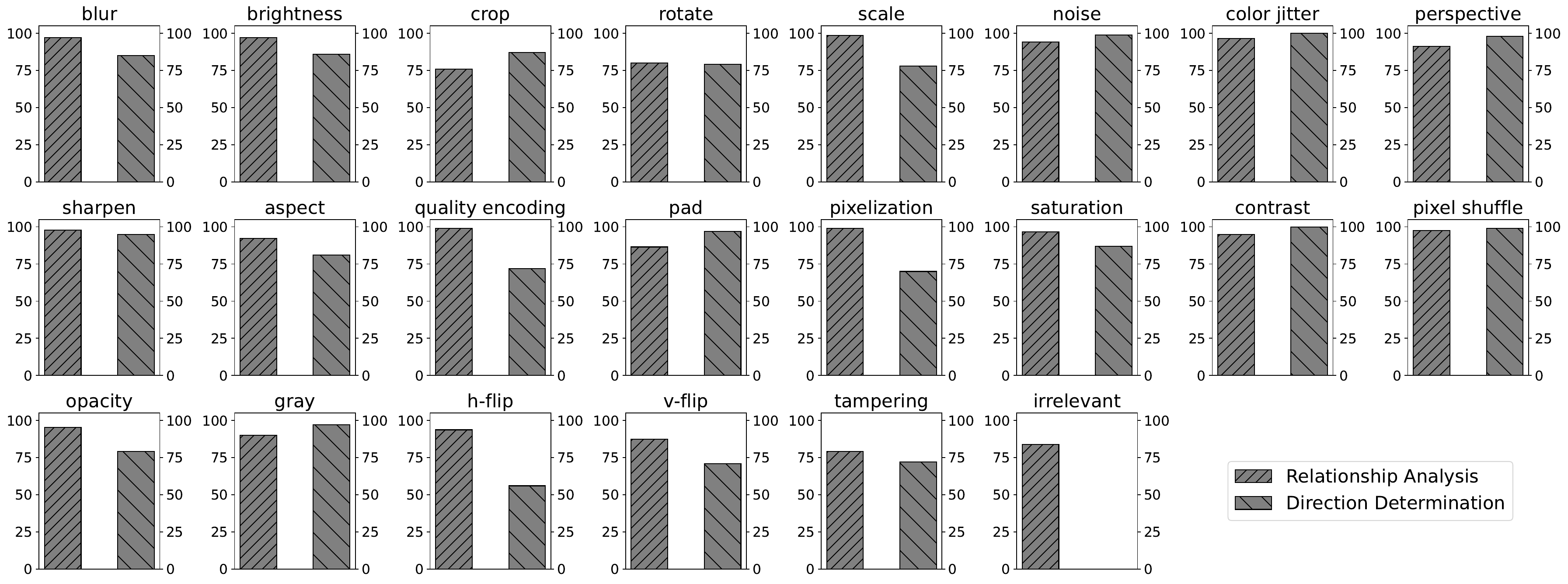}
  \caption{
The evaluation results of our models in 22 different scenarios are presented. These include the tasks of MR analysis and direction determination. We only conducted an evaluation of the image MR identification task in unrelated scenarios.
  }
  \label{fig:type_box}
\end{figure*}

The MR analytical model demonstrates robust performance across diverse modified images, achieving an average accuracy of 91.97\%.
The proposed model demonstrates exceptional resilience to color-space perturbations, maintaining over 95.0\% accuracy in scenarios involving color jitter, contrast adjustments, brightness variations, saturation shifts, and grayscale-conversion operations. 
Under geometric transformations, such as perspective warping, aspect ratio adjustments, padding operations, scale variations, horizontal/vertical flipping, and rotational operations, the model achieves consistently high accuracy (exceeding 90.0\%) across all test cases. 
For pixel-level distortions that only modify pixel values without changing color or structure, including pixelization, quality encoding, Gaussian noise injection, sharpening, opacity adjustment, and shuffling, the model maintains robust performance with an average accuracy of 97.1\%. 
Even for semantic content modifications (cropping and tampering), our model reaches 83.8\% accuracy in complex tampering scenarios, demonstrating functional performance.
Furthermore, our model effectively filters out irrelevant images, consistent with its previously validated robustness in constructing provenance graphs.

The direction determination model demonstrates considerable robustness, achieving an average accuracy of 85.3\% across 21 modification scenarios.
The model demonstrates exceptional performance in color-space modifications with an average accuracy of 92.0\%, including perfect performance in color jitter and contrast adjustments, while maintaining strong results for brightness variations and saturation shifts. 
However, it exhibits higher variability when facing structural modifications. While our model achieves perfect performance in perspective warping and padding operations, it shows degradation in processing images that have undergone aspect ratio adjustment, scaling and rotation operations, especially horizontal flipping, attributed to disrupted spatial coherence in JPEG macroblock structures. 
Regarding pixel-level distortions, our model demonstrates strong robustness in Gaussian noise injection, sharpening and shuffling operations (more than 95.0\%). However, it struggles with remaining operations, which may destroy initial JPEG compression artifacts within images.
Finally, under semantic manipulations, the model shows reliable direction determination performance, achieving 87.0\% in cropping operation and 72.0\% in tampering scenario.

\section{Conclusion}
Image provenance analysis plays a crucial role in digital image forensics.
To address the limitations of current research and overcome the challenges of image provenance analysis, this paper conducts several explorations regarding provenance accuracy and computational overhead. We design a scalable pipeline for large-scale images, incorporating the proposed filtering method and directed graph construction methods. Our method enhances provenance filtering through MR tracing, which allows us to comprehensively discover images sharing MR with the query images. By local features matching to mine the relationships
among images and capturing compression artifacts to determine modification direction, our method is able to establish directed edges between image pairs that share direct MR with precision, and handle diverse modified images with robustness, thus accurately constructing directed provenance graphs. Moreover, by optimizing computational costs, we enhance the scalability of our pipeline, enabling it to operate with time computational complexity.  Comprehensive experiments demonstrate significant improvements in both accuracy and efficiency across various tasks. In directed provenance graph construction, our method achieves a 16.7-56.1\%  improvement in VEO$^{\diamond}$ score in end-to-end provenance analysis. Importantly, our method requires only an average of 3.0 seconds per image at scale (10 million images), which is far below the SOTA's 12-minute processing time.


\bibliographystyle{ACM-Reference-Format}
\bibliography{ref}


\end{document}